\documentclass[runningheads]{llncs}
\usepackage[T1]{fontenc}
\usepackage{graphicx,verbatim}
\usepackage{xcolor}
\usepackage{dirtytalk}
\usepackage{amssymb}
\usepackage{amsmath}
\usepackage{hyperref}
\usepackage{algpseudocode}
\usepackage{subcaption}
\usepackage{cite}
\usepackage{tabularx}
\usepackage{arydshln}
\begin{document}
\title{Clinically-guided Data Synthesis for Laryngeal Lesion Detection}
%
\author{Chiara Baldini\inst{1,2}
\and
Kaisar Kushibar\inst{3} \and
Richard Osuala\inst{3} \and Simone Balocco\inst{3,4} \and Oliver Diaz\inst{3,4} \and Karim Lekadir\inst{3,5} \and Leonardo S. Mattos\inst{1}}
\authorrunning{C. Baldini et al.}
%
\institute{Biomedical Robotics Laboratory, Department of Advanced Robotics, Istituto Italiano di Tecnologia, Genoa, Italy \\ \email{chiara.baldini@iit.it} 
\and Department of Informatics, Bioengineering, Robotics, Systems Engineering, University of Genova, Genoa, Italy \and Departament de Matem\`atiques i Inform\`atica, Universitat de Barcelona, Spain 
\and Computer Vision Center, Bellaterra, Spain \and
Instituci\`o Catalana de Recerca i Estudis Avan\c cats (ICREA), Barcelona, Spain }



\maketitle              
\begin{abstract}
\setcounter{footnote}{0}
Although computer-aided diagnosis (CADx) and detection (CADe) systems have made significant progress in various medical domains, their application is still limited in specialized fields such as otorhinolaryngology. In the latter, current assessment methods heavily depend on operator expertise, and the high heterogeneity of lesions complicates diagnosis, with biopsy persisting as the gold standard despite its substantial costs and risks. A critical bottleneck for specialized endoscopic CADx/e systems is the lack of well-annotated datasets with sufficient variability for real-world generalization. This study introduces a novel approach that exploits a Latent Diffusion Model (LDM) coupled with a ControlNet adapter to generate laryngeal endoscopic image-annotation pairs, guided by clinical observations. The method addresses data scarcity by conditioning the diffusion process to produce realistic, high-quality, and clinically relevant image features that capture diverse anatomical conditions. The proposed approach can be leveraged to expand training datasets for CADx/e models, empowering the assessment process in laryngology. Indeed, during a downstream task of detection, the addition of only 10\% synthetic data improved the detection rate of laryngeal lesions by 9\% when the model was internally tested and 22.1\% on out-of-domain external data. Additionally, the realism of the generated images was evaluated by asking 5 expert otorhinolaryngologists with varying expertise to rate their confidence in distinguishing synthetic from real images. This work has the potential to accelerate the development of automated tools for laryngeal disease diagnosis, offering a solution to data scarcity and demonstrating the applicability of synthetic data in real-world scenarios. We publicly share our codebase at \url{https://github.com/ChiaraBaldini/endoLDMC.git}.


\end{abstract}
\section{Introduction}

Medical imaging is evolving due to the emergence of Computer-aided Diagnosis and Detection (CADx/e) systems, which enable more accurate diagnoses by using deep-learning algorithms to assist clinicians in identifying pathological patterns in medical images.
Domains such as radiology, oncology, and general endoscopy benefit from large well-annotated datasets, driving significant improvements\cite{harris2019systematic, guo2022review, wen2022characteristics, zhu2023public}. 
However, in specialized domains like laryngology, the application of such systems remains limited, primarily due to the lack of annotated data. While some public laryngology datasets exist, their annotations are often tailored to specific, narrow tasks, such as evaluating video frame quality \cite{Moccia2018-ku}, analyzing vocal cord mobility \cite{Andrade-Miranda2024-tf, Low2021-er}, performing semantic segmentation \cite{laves2019dataset}, or classifying lesions \cite{YIN2021207}. Moreover, these datasets are typically acquired from a single clinical center and fail to capture the full spectrum of possible laryngeal lesions. This insufficient diversity in data representation and annotation severely limits their applicability in real-world clinical settings, where broader anatomical and pathological variability is required to address complex diagnostic tasks.
Although the availability of annotated datasets remains limited, the need for CADx/e tools in laryngology is becoming more evident. This demand is driven by the operator-dependent nature of current diagnostic methods and the inconsistency introduced by subjective assessments. While techniques such as Narrow Band Imaging (NBI) \cite{watanabe2009value} have been introduced to improve the visualization of neoangiogenic changes, their effective use relies on sophisticated endoscopic systems and significant operator expertise. Recently published studies have attempted to develop algorithms for detecting, classifying, and segmenting laryngeal lesions \cite{sampieri2023artificial}. Nevertheless, only a small portion of them have tested model reliability and robustness on external populations and validated the applicability in the clinical environment due to diverse data scarcity.

Synthetic data generation techniques have become increasingly used by researchers to augment existing datasets and overcome issues related to variability limitations. Recent advances in generative models, such as Generative Adversarial Networks (GANs) and Variational Autoencoders (VAEs), have demonstrated remarkable success in producing realistic synthetic images for medical domains, including general endoscopy \cite{li2024endora, yoon2022colonoscopic, bajhaiya2023deep, liu2025polyp, diamantis2022endovae, martyniak2024simuscope, osuala2023medigan}. To the best of our knowledge, no studies have explored the synthesis of images for laryngeal lesion diagnosis. The complex anatomical and pathological characteristics of the laryngeal region make it a particularly challenging area for synthetic approaches, requiring advanced generative models to produce clinically relevant images that capture its inherent variability. One approach to ensuring the clinical relevance of synthetic data is to condition the generation process on clinical observations, such as acquisition parameters or spatial information of the lesion. Furthermore, how to maximize the potential of synthetic data by optimally selecting them to improve downstream task performance is rarely explored.

This study addresses the previously identified challenges through the following contributions:
\begin{enumerate}
    \item Generation of synthetic laryngeal endoscopic image-annotation pairs by leveraging Latent Diffusion Models (LDMs) conditioning by clinical observations.
    \item 
    Proposing a synthetic data selection mechanism to improve the performance of a downstream lesion detection task.
    \item Validation of the realism of the generated images through a human-observer study. 
    
\end{enumerate}

\section{Methods}
\subsection{Laryngeal datasets}
Laryngeal endoscopic images of different resolutions were collected independently from two hospitals\footnote{For each center, the local Institutional Ethics Committee approval was obtained.} (2014–2023) during routine practice, using flexible or rigid endoscopes\footnote{Olympus Medical System Corporation, Tokyo, Japan} in White-Light (WL) and NBI modes. The first medical center involved in the data collection was the Otolaryngology-Head and Neck Unit of the Oncologic Hospital Saint Savvas (Athens, Greece), providing a total amount of 909 laryngeal images, termed as \textit{internal} in the following as they were used for training and internal validation of the LDM and the subsequent downstream task. The second center was the Otolaryngology-Head and Neck unit of IRCSS Hospital San Martino (Genova, Italy), where 88 images were extracted from video laryngoscopies. This \textit{external} dataset was considered only for the validation of the downstream task. Each image was labeled with a bounding box that encompassed the surface of each lesion and a class associated with the type of lesion – cyst, granuloma, leukoplakia, polyp, papilloma, Reinke’s edema, squamous cell carcinoma –, according to histopathological confirmation or expert consensus. Such annotations were carried out by expert otorhinolaryngologists at each center and revised by two other experts at the IRCSS Hospital San Martino.

\subsection{Clinically-guided Latent Diffusion Model}
\begin{figure}
    \centering
    \includegraphics[width=0.95\linewidth]{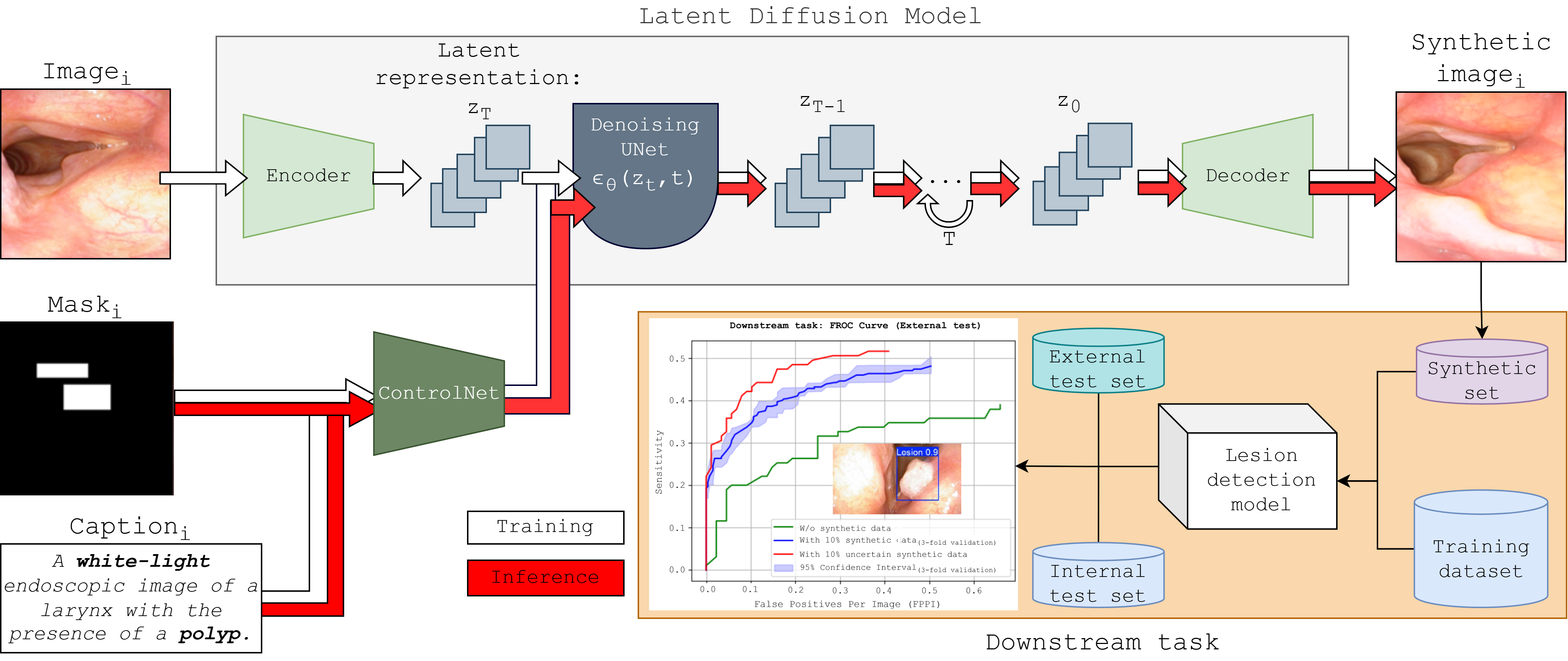}
    \caption{ Overview of the proposed method. During training (white arrow path), a caption with information on the optical modality, the lesion type, and the mask of the annotated bounding boxes were used as inputs to control the learning process of a Latent Diffusion Model (LDM). Realistic synthetic images were generated in the inference step (red arrow path), under the guidance of captions and masks, to enhance the performance of a downstream detection task.}
    \label{fig:method_overview}
\end{figure}

The proposed method generates synthetic images using a Latent Diffusion
Model (LDM) for realistic reconstruction and a ControlNet module to preserve
clinical relevance.

\subsubsection{Latent Diffusion Model}
A Diffusion Model (DM) progressively perturbs the observed data distribution, \math{x_{0}}\endmath, by gradually adding Gaussian noise (\math{\epsilon \sim \mathcal{N}(0,\,1)}\endmath) until it converges to a given prior \math{x_{t}}\endmath. This process is known as diffusion. It then learns a reverse diffusion process (\math{\epsilon_{\theta}}\endmath) to reconstruct and restore the original data, starting from this prior distribution and gradually produces less noisy samples \math{x_{t-1}}\endmath, \math{x_{t-2}}\endmath, …, until reaching the final sample \math{x_{0}}\endmath. The loss function can be written as: 
\begin{equation}
    L_{DM}(\theta) = \mathbb{E}_{x_{0}, \epsilon \sim \mathcal{N}(0, 1), t \sim U[1,T]} \left[ \lVert \epsilon - \epsilon_{\theta}(x_{t}, t) \rVert^2_{2} \right].
\end{equation}

The core concept of LDMs is to apply the diffuse and reverse diffusion processes in the low dimensional latent feature space of the inputs (\math{z_{t}}\endmath), resulting in reduced complexity \cite{Rombach2021HighResolutionIS}. The input image (\math{x_{0}}\endmath) is fed through an encoder, that outputs a latent vector \math{z_{0}}\endmath. For T iterations, a diffusion process adds noise to \math{z_{t}}\endmath, and then a denoising process is carried out through a UNet denoiser model. In the last part, a decoder converts the restored latent signal (\math{z_{0}}\endmath) back to a clean signal in the original image space. The loss can be modified as follows: 
\begin{equation}
    L_{LDM}(\theta) = \mathbb{E}_{z_{0}, \epsilon \sim \mathcal{N}(0, 1), t \sim U[1,T]} \left[ \lVert \epsilon - \epsilon_{\theta}(z_{t}, t) \rVert^2_{2} \right].
\end{equation}
In this work, pre-trained autoencoder and text encoder models\cite{Rombach2021HighResolutionIS} were used
following a transfer learning strategy to speed up the training process of the
whole LDM.

\subsubsection{ControlNet}
ControlNet\cite{zhang2023adding} is an extended module that can be added to diffusion models to enable fine-grained control over the image generation process
by incorporating additional conditioning inputs. The integration of extra information — such as edge, depth, or segmentation masks - allows the model to
generate data that maintains the overall realism provided by the latent diffusion
model but respects closely the structural or semantic indications of the conditioning signal. This level of control is even more critical in the medical field, especially when replicating pathological conditions, where the structural and
spatial characteristics of abnormalities are key factors for clinical interpretation.
We integrated detailed observations derived from clinical practice into ControlNet, learning patterns between laryngeal lesions and expert annotations, while
also parsing biopsy results. Hence, our approach allows the creation of synthetic
images that contain clinically relevant features. In particular, we decided to pass
two different inputs to the ControlNet module. An image caption containing
information about the optical modality used for the acquisition – WL or NBI
– and the class of the laryngeal lesion annotated by our clinical partners was
provided as the first control condition. An example of a caption is shown in Fig.
\ref{fig:method_overview}: \say{A white-light endoscopic image of a larynx with the presence of a polyp}.
In addition, a mask of the annotated bounding boxes was included in the conditioning process to provide information about the location of lesions. During inference, the model takes random noise as input and generates a new synthetic
endoscopic image of a larynx, having the characteristics of the optical modality
specified in the provided caption, with the specific type of lesion located in the
desired area of the image.

\subsubsection{Implementation details}
To train the LDM model with ControlNet, the \textit{internal} dataset was split into three subsets with a ratio of 80:10:10, resulting in 727 images for training, 91 for validation, and 91 for testing. Initially, the LDM
model was fine-tuned for 100 epochs, with a batch size of 4, an initial learning rate of 0.00005, and an image resolution of 640 $\times$ 640. Then, a 3-channel mask of the bounding box was obtained for each image, starting from the coordinates annotated by experts in YOLO format. A caption was also generated as a textual prompt for each image considering the lesion class and modality information as previously explained. Hence, the LDM model with the ControlNet module was further fine-tuned, with the same parameters as the previous step, to integrate additional features encoded by the ControlNet. To test the generalization ability of the LDM, 727 new masks were created via random rotations and scaling operations of those of the \textit{internal} dataset, and were utilized, together with the corresponding captions, to generate a larger set of diversified synthetic images, using ControlNet conditioning and guidance scales empirically set to 1.0. We ran all the experiments using an NVIDIA A100 GPU with 80GB of memory.

\subsection{Evaluation metrics}
Our goal is to generate diverse, clinically relevant synthetic images distinct from the originals to enrich CADx/e training datasets. Hence, an evaluation metric that computes the similarity between real and generated images, such as the Fréchet Inception Distance (FID) \cite{NIPS2017_8a1d6947}, does not fully align with our objectives and may undervalue the contribution of clinically diverse synthetic data. For those reasons, we evaluated the proposed method with the \textit{FIDratio} \cite{osuala2023medigan}, considering the FID between real and synthetic samples (\textit{FID\textsubscript{rs}}) related to the FID between two sets of real samples (\textit {FID\textsubscript{rr}}), which produced a high FID value due to high diversity. It was calculated as \math{FID_{ratio} = 1- (FID_{rs}-FID_{rr})/FID_{rs}}\endmath.  
The Inception Score (IS) \cite{salimans2016improved} was also computed to evaluate data diversity compared to real images. Moreover, two other approaches were taken into account to analyze the advantages of the synthetic images and validate their quality, as detailed in the following.

\subsubsection{Downstream task}
We explored the practical utility of the generated synthetic data by augmenting the available \textit{internal} laryngeal dataset. Specifically, we combined the synthetic images with the real ones to train a lesion detection model 
that can learn to handle a wider variety of cases, improving its robustness. 
As lesion detector, we utilized the pre-trained YOLOv8 nano model from the Ultralytics GitHub repository\footnote{https://github.com/ultralytics}. This model is identical to the one currently in clinical use in a hospital that collaborates with our team, and was chosen due to its capability of working efficiently in real-time. We trained the lesion detector for 100 epochs, first only with the real images of the \textit{internal} dataset and then gradually added 5\%, 10\%, 20\%, 40\%, and 80\% synthetic samples. 
Taking into account the percentage that achieved the best results, we carried out a 3-fold cross-validation by repeating the detection learning process with 3 sets of different synthetic data. However, as demonstrated by the cross-validation experiment, the random selection of synthetic samples can slightly impact performance variability. To address this issue, we defined an Uncertainty Estimation (UE)
strategy for extending the training dataset with “challenging samples” based on
the uncertainty of detection predictions. Considering the 3 models obtained from
cross-validation, all generated synthetic data were predicted, and the variance
between the confidence scores returned by each detection model was calculated
for each image. Finally, only the top 10\% of samples were selected after sorting
the elements in descending order of variance.
For all the experiments, we applied the AdamW optimizer, an early-stopping strategy of 50 epochs, a batch size of 32, an initial learning rate of 0.001, and the default data augmentation. The detection performance was tested - in terms of precision, recall, and Average
Precision\textsubscript{@IoU=0.5} (AP\textsubscript{@IoU=0.5}) -  both on the \textit{internal} test set and on the unseen \textit{external} data.

\subsubsection{Human-observer study}
We conducted a clinical investigation about the quality of the images by selecting a subset of 20 images, 10 real and 10 synthetic, and asking 5 of our clinical collaborators, with 7, 6, 11, 6, and 8 years of experience, to blindly rate the realism on a \say{Likert} scale from \say{strongly disagree} to \say{strongly agree} concerning the sentence \say{\textit{This is a REAL image}}. We assigned probability values to each possible vote, with the aim of computing the Area Under the ROC Curve (AUC) of each user and evaluating their accuracy in correctly identifying real samples. In particular, the \say{Likert} scale - [Strongly disagree, Disagree, Slightly disagree, Slightly agree, Agree, Strongly agree] - was translated into the following probability values: [0.05, 0.23, 0.41, 0.50, 0.77, 0.95]. 

\section{Results and Discussion}
\subsection{Generating images and annotations pairs}
\begin{figure}
    \centering
    \includegraphics[width=\linewidth]{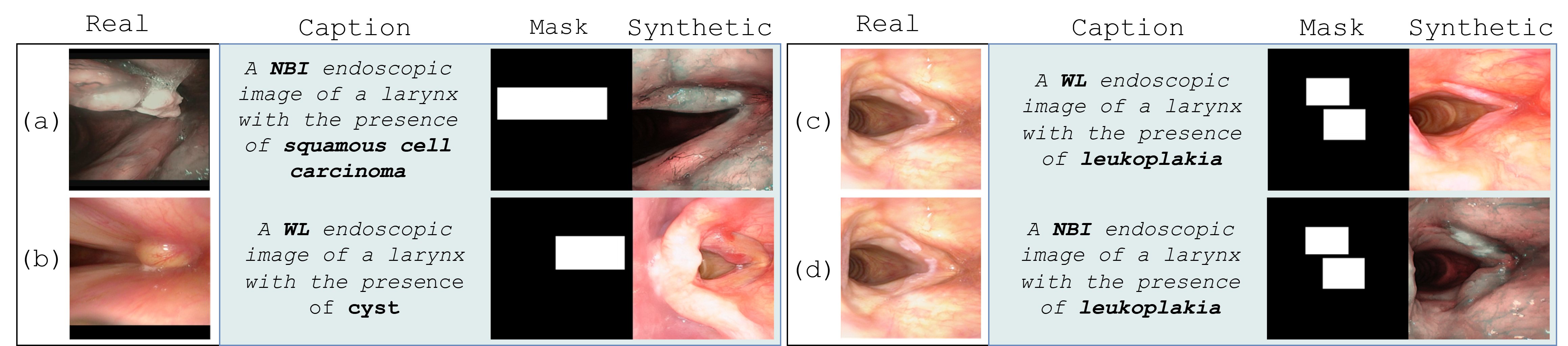}
    \caption{Starting from real images (1st column), a caption containing relevant information (2nd column) and the manual-annotated mask of the lesion bounding boxes (3rd column) were used to generate new samples (4th column).}
    \label{fig:qualitative_results}
\end{figure}

Fig. \ref{fig:qualitative_results} presents qualitative results of the image generation process, highlighting the ability of our method to produce realistic laryngeal images. The generated data reflect variations in vocal cord anatomy, including different lesion types -- squamous cell carcinoma, cyst, leukoplakia -- and acquisition settings -- WL, NBI --, emphasizing the model's capacity to reconstruct complex and diverse patterns. As shown in cases (c) and (d) of Fig. \ref{fig:qualitative_results}, we observed that the same mask can be associated with diverse synthetic outputs by using different captions. The \textit{FIDratio} metric was 0.836 as the difference between the \textit{FID\textsubscript{rs}} and \textit{FID\textsubscript{rr}} was small, while the IS was 3.840 ± 0.270 for real samples and 4.240 ± 0.270 for synthetic data, with both metrics confirming a high variety of the synthetic data.

\subsection{Improving detection by adding synthetic images}


\begin{table}[b]
\caption{Analysis of the impact of adding 10\% synthetic data, either randomly selected or based on the Uncertainty Estimation (UE) strategy.}
\label{3fold_table}
\resizebox{\textwidth}{!}{
\centering
\renewcommand{\arraystretch}{1.3} 
\begin{tabular}{|l|c|c|c|c|c|c|} \hline  
\textit{Train dataset} & \multicolumn{3}{|c|}{\textit{Internal test}} & \multicolumn{3}{|c|}{\textit{External test}} \\ \hline  
 & \textbf{Precision} & \textbf{Recall} & \textbf{AP\textsubscript{\tiny{@IoU=50}}} & \textbf{Precision} & \textbf{Recall} & \textbf{AP\textsubscript{\tiny{@IoU=50}}} \\ \hline
 \textit{Real} & 0.761 & 0.763 & 0.798 & 0.498 & 0.376 & 0.359 \\ \hline
\textit{Real+10\% synthetic} & & & & & & \\ 
Random selection (fold 1)& 0.848 & 0.851 & 0.895 & 0.614 & 0.526 & 0.526 \\ 
Random selection (fold 2)& 0.828 & 0.889 & 0.900 & 0.660 & 0.463 & 0.518 \\
Random selection (fold 3)& 0.809 & 0.789 & 0.846 & 0.631 & 0.468 & 0.515 \\ \hdashline 
Mean\textsubscript{$\pm$std} & 0.828\textsubscript{$\pm$0.016} & \textbf{0.843}\textsubscript{$\pm$0.041} & 0.880\textsubscript{$\pm$0.024} & 0.635\textsubscript{$\pm$0.019} & \textbf{0.485}\textsubscript{$\pm$0.028} & 0.519\textsubscript{$\pm$0.004} \\ 
\textbf{Top uncertain selected} & \textbf{0.905} & 0.833 & \textbf{0.888} & \textbf{0.800} & 0.464 & \textbf{0.580} \\ \hline
\end{tabular}}
\end{table}


In Fig. \ref{fig:plateaus}, we present the results of the downstream detection task. When testing on both the \textit{internal} and \textit{external} test sets, the introduction of synthetic data resulted in average performance improvements. 
On the \textit{internal} test set, AP\textsubscript{@IoU=0.5} peaked at 0.895 with 10\% synthetic data, after which the performance reached a plateau.
Similarly, the \textit{external} test indicated that the best
performance was achieved with an addition of 10\% synthetic data.
The advantage of adding the 10\% of randomly selected synthetic data was further confirmed by the results of the 3-fold cross-validation (Table \ref{3fold_table}), with an average AP\textsubscript{@IoU=0.5} increment regarding the use of only real data of +8.2\% and +16.0\% for \textit{internal} and \textit{external} tests, respectively. Nevertheless, the highest performance was observed with the uncertainty-based selection strategy, i.e. +9.0\% and +22.1\% for \textit{internal} and \textit{external} AP\textsubscript{@IoU=0.5}, respectively, meaning that extending the existing dataset with \say{challenging samples} can substantially improve the model's ability to detect hard and diverse lesions and to generalize on unseen images. 

\subsection{Real versus synthetic: human-observer study}

\begin{figure}
    \centering
    \begin{subfigure}[b]{0.42\linewidth}
    \includegraphics[width=\textwidth]{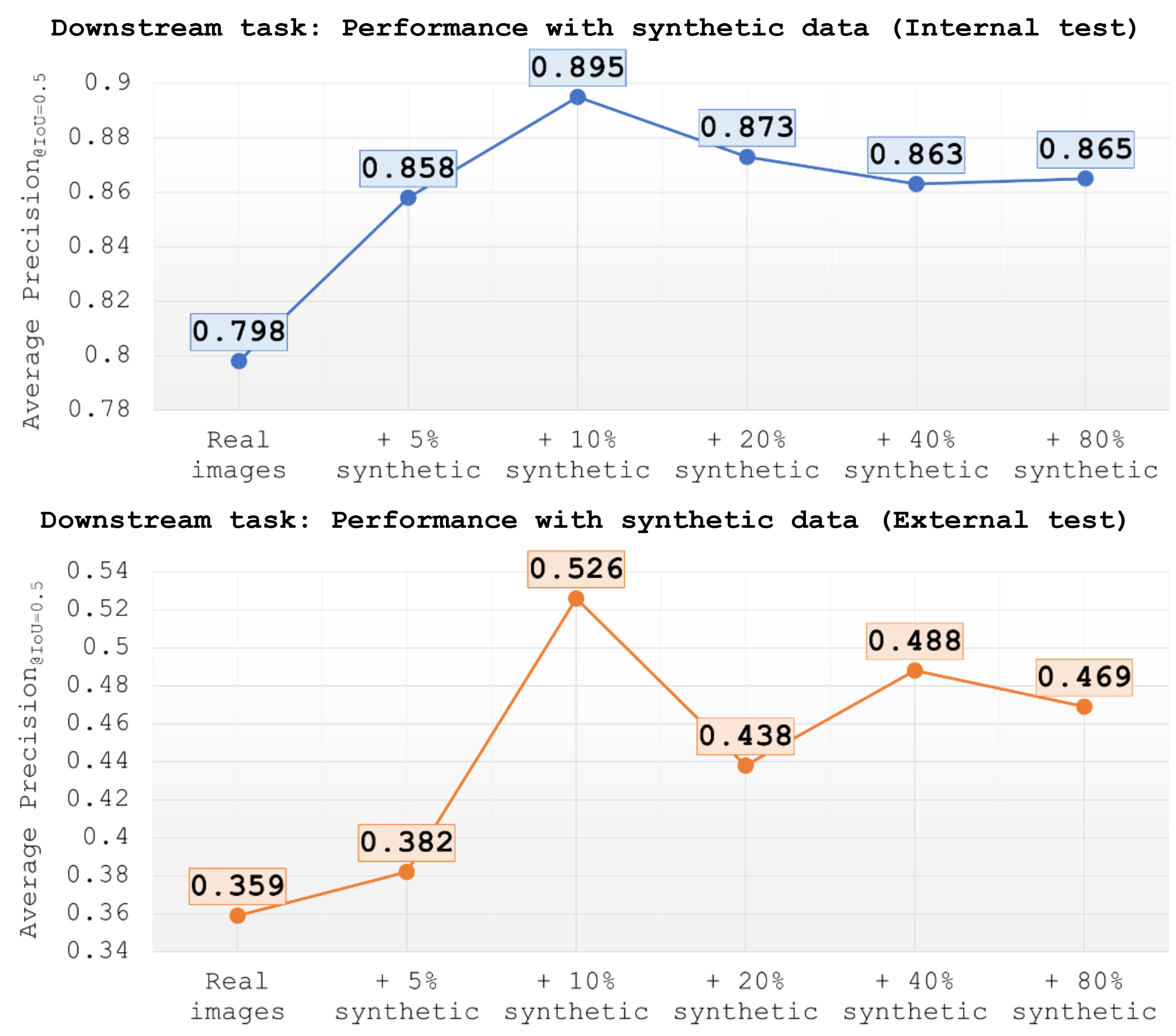}
    \caption{}
    \label{fig:plateaus}
    \end{subfigure}%
    \begin{subfigure}[b]{0.58\linewidth}
    \includegraphics[width=\linewidth]{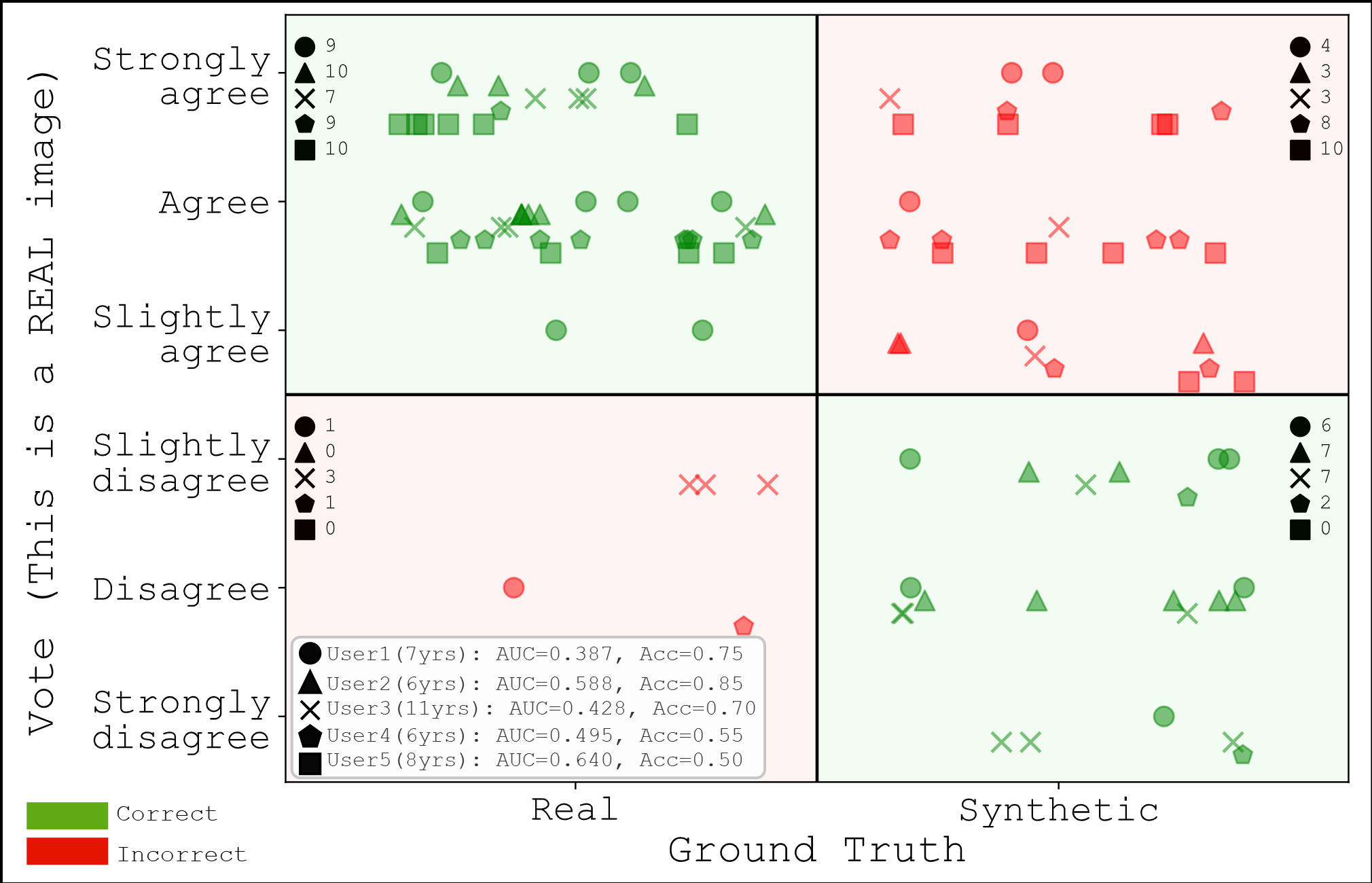}
    \caption{}
    \label{fig:clinical_investigation}
    \end{subfigure}
    \caption{
    \textbf{(a)} Evolution of the Average Precision\textsubscript{@IoU=0.5} by gradually adding synthetic images to the training dataset and testing the performance both internally and externally, showing peaks at + 10\% synthetic samples.  \textbf{(b)} Votes from otorhinolaryngologists assessing the realism of images, with confidence levels represented on the vertical axis and marker colors indicating whether the classification of images as real or synthetic was correct.}
\end{figure}

In Figure \ref{fig:clinical_investigation}, the responses of the enrolled otorhinolaryngologists were displayed in green for each correct real or synthetic image identified and red for incorrect answers. On the y-axis, the predictions were discriminated by the confidence level of the user's answer. Among all users, the AUC and accuracy values were 0.507$\pm$0.095 and 0.670$\pm$0.129, respectively. Indeed, we can empirically affirm the realism of the synthetic data as clinicians struggled to recognize synthetic samples, often classifying them as 
real cases.

\section{Conclusion}
This study introduced an LDM and ControlNet-based framework for generating synthetic laryngeal image-annotation pairs, pioneering data generation in laryngology. We showed that the addition of challenging synthetic samples determined via uncertainty estimation improves lesion detection, and successfully validated the clinical realism of generated samples through expert reviews. Future work will further enhance generation diversity, explore lesion-specific synthesis in privacy-preserving settings, and evaluate the data’s utility in other tasks like lesion
classification.



\bibliographystyle{splncs04}
\bibliography{Paper-4594}
\end{document}